\documentclass[11pt]{article}
\usepackage{moriond,epsfig}

\def\Journal#1#2#3#4{{#1} {\bf #2}, #3 (#4)}

\def\PLB{{\em Phys. Lett.}  B}

\def\PRD{{\em Phys. Rev.} D}
\def\NPA{{\em Nucl. Phys.} A}

\def\JHEP{\em JHEP}

\def\be{\begin{equation}}
\def\ee{\end{equation}}
\def\bea{\begin{eqnarray}}
\def\eea{\end{eqnarray}}

\def\lsim{\raise0.3ex\hbox{$<$\kern-0.75em\raise-1.1ex\hbox{$\sim$}}}
\def\gsim{\raise0.3ex\hbox{$>$\kern-0.75em\raise-1.1ex\hbox{$\sim$}}}

\begin{document}
\vspace*{4cm}
\title{The QCD equation of state from improved staggered fermions}

\author{ C.~Schmidt (for hotQCD Collaboration) }

\address{Universit\"at Bielefeld, Fakult\"at f\"ur Physik, Universit\"atsstr. 25, \\
D-33615 Bielefeld, Germany}

\maketitle\abstracts{
We calculate the equation of state in 2+1 flavor QCD at finite
temperature with physical strange quark mass and almost physical light
quark masses using lattices with temporal extent
$N_{\tau}=8$. Calculations have been performed with two different
improved staggered fermion actions, the asqtad and p4 actions.
Overall, we find good agreement between results obtained with these
two $O(a^2)$ improved staggered fermion discretization schemes.  A
comparison with earlier calculations on coarser lattices is performed
to quantify systematic errors in current studies of the equation of
state.  We also present results for observables that are sensitive to
deconfining and chiral aspects of the QCD transition on $N_\tau=6$ and
$8$ lattices. We find that deconfinement and chiral symmetry
restoration happen in the same narrow temperature interval.}

\section{Introduction}
A detailed and comprehensive understanding of the thermodynamics of
quarks and gluons, e.g. of the equation of state is most desirable and
of particular importance for the phenomenology of relativistic heavy
ion collisions. In
particular, the interpretation of recent results from RHIC on jet
quenching, hydrodynamic flow, and charmonium production~\cite{RHIC}
rely on an accurate determination of the energy density and pressure
as well as an understanding of both the deconfinement and chiral
transitions.  For vanishing chemical potential, which is appropriate
for experiments at RHIC and LHC, lattice calculations of the EoS~\cite{aoki,milc_eos,RBCBi_eos,hotQCDeos} 
as well as the transition temperature~\cite{Tc} can be performed with an almost realistic
quark mass spectrum. In addition, calculations at different values of
the lattice cutoff allow for a systematic analysis of discretization
errors and will soon lead to a controlled continuum extrapolation of
the EoS with physical quark masses.

\section{Improved actions and the calculational setup}
Studies of the QCD equation of state are most advanced in lattice
regularization schemes that use staggered fermions. In this case,
improved actions have been developed that reduce ${\cal O}(a^2)$
discretization effects efficiently. Such a reduction is mandatory
because a quantitative lattice determination of the EoS requires
rather coarse lattices as one has to
take into account that pressure and energy density are dimension 4
operators, which are difficult to calculate on fine lattices.  
At tree level which is relevant for the high
temperature phase both the asqtad and the p4 actions were found to
give rise to only small deviations from the asymptotic ideal gas limit
already on lattices with temporal extent $N_\tau = 6$. At $N_\tau =8$
the differences from the continuum Stefan-Boltzmann value are at the 1
\% level~\cite{heller}.  At moderate values of the temperature
non-perturbative effects contribute to the cutoff dependence. In
particular in the hadronic phase at low temperature, the breaking of
flavor (taste) symmetry, inherent to staggered fermions away from the
continuum limit, leads to an ${\cal O}(a^2)$ distortion of the
pseudoscalar hadron spectrum and may influence the
thermodynamics in the confined phase.  In order to judge the
importance of different effects that contribute to the cutoff
dependence of thermodynamic observables, we have performed
calculations with both p4 and asqtad actions which deal with these
systematic effects in different ways.  We extend previous lattice
calculations on $N_\tau=6$ lattices (asqtad~\cite{milc_eos} and p4~\cite{RBCBi_eos}) 
to $N_\tau=8$ lattices~\cite{hotQCDeos}.  In
these studies the bare strange quark mass ($m_s$) was tuned such
that the zero temperature kaon mass acquired a constant value of 500
MeV (p4) and a slightly larger value for asqtad (570~MeV).  The
light quark mass was hold at a fixed ratio to the strange one of $m_l
= 0.1m_s$ leading to a Goldstone pion mass of about 220~MeV at $T=0$.
The temperature scale was set by using the Sommer parameter $r_0=0.469(7)$~fm,~\cite{gray} 
which is determined from the shape of the heavy quark potential.
%For a reliable determination of the EoS to be useful in
%phenomenology and for the interpretation of experimental results,
%it is therefore important to carefully control discretization effects
%at low as well as at high temperature and perform the continuum limit.

\section{The trace anomaly, pressure and energy density}
The basic thermodynamic quantity most convenient to calculate on the lattice is 
the trace anomaly in units of the fourth power of the
temperature $\Theta^{\mu\mu}/T^4$. This is given by the derivative
of $p/T^4$ with respect to the temperature,
\begin{eqnarray}
\frac{\Theta^{\mu\mu} (T)}{T^4} \equiv \frac{\epsilon - 3p }{T^4}  =  
T \frac{\partial}{\partial T} (p/T^4)~.
\label{delta}
\end{eqnarray}
As the pressure is given by the logarithm of the partition function, 
$p/T = V^{-1} \ln Z$,
the calculation of the trace anomaly requires only the evaluation of rather simple
expectation values.
\begin{figure}
\begin{center}
\resizebox{0.475\textwidth}{!}{%
  \includegraphics{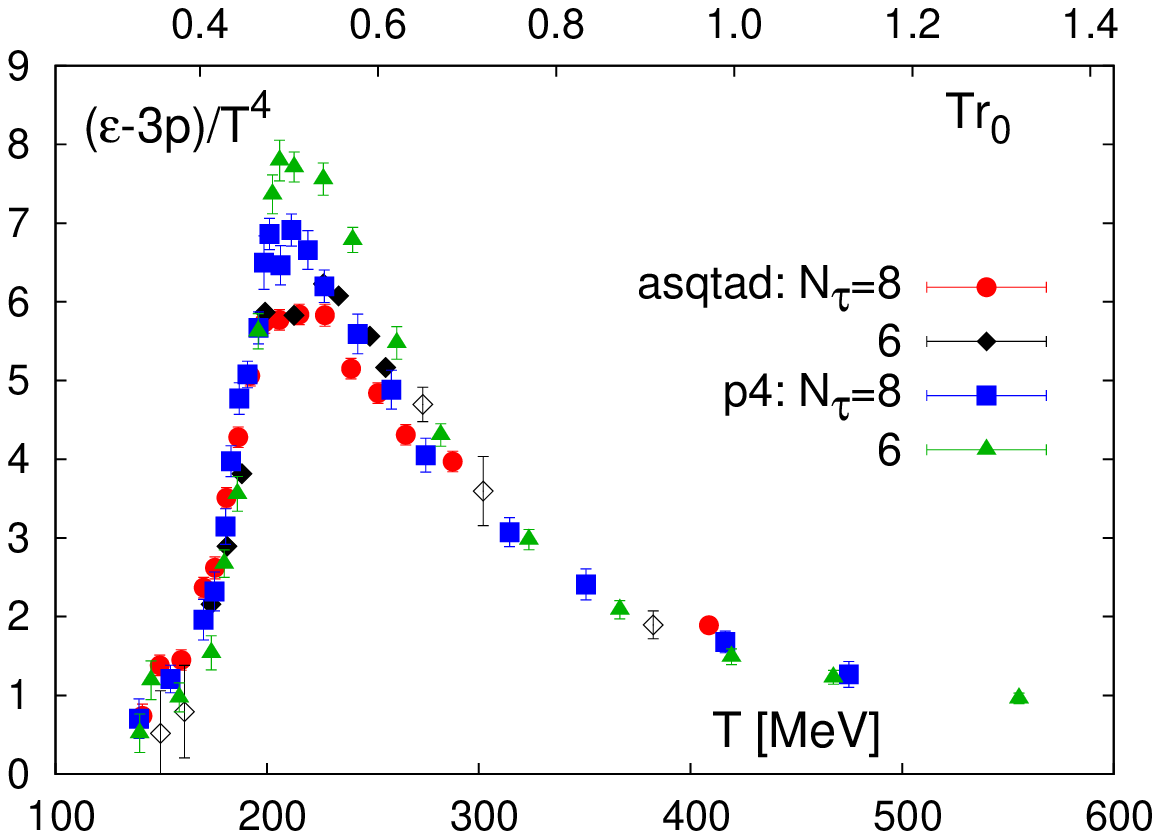}
}
\resizebox{0.3304\textwidth}{!}{%
  \includegraphics{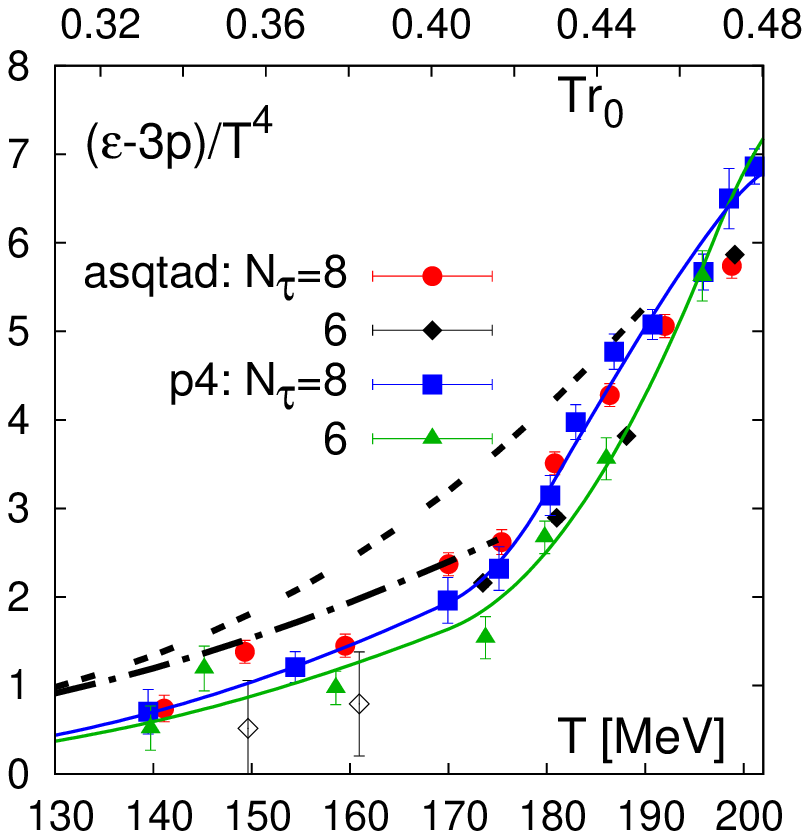}
}
\end{center}
\caption{The trace anomaly, $(\epsilon -3p)/T^4$ on $N_\tau=6$ and $8$
lattices. Results are obtained with the p4 and asqtad actions. 
The right panel shows the low temperature part of the trace anomaly in detail. 
Here we show fits to the data, as well as a comparrison with 
the resonance gas model (dashed and dash-doted lines).\label{fig:e-3p}}
\end{figure}
In Fig.~\ref{fig:e-3p}, we show results for $\Theta^{\mu\mu}/T^4$
obtained with both the asqtad and p4 actions. The new $N_\tau=8$ results~\cite{hotQCDeos}
have been obtained on lattices of size $32^3 \times 8$ and the additional
zero temperature calculations, needed to carry out the necessary vacuum
subtractions, have been performed on $32^4$ lattices. The $N_\tau=6$ results 
are taken from Ref.~3 (asqtad) and Ref.~4 (p4), respectively.

We find that the results with asqtad and p4 formulations are in good
agreement. In particular, both actions yield consistent results in the
low temperature range (see Fig~\ref{fig:e-3p} (right)), in which
$\Theta^{\mu\mu}/T^4$ rises rapidly, and at high temperature, $T \gsim
300$~MeV. This is also the case for the cutoff dependence in these two
regimes. At intermediate temperatures, $200\;{\rm MeV}\lsim T\lsim
300\;{\rm MeV}$, the two actions show differences and cutoff effects are
more pronounced.
%   The maximum in
%$\Theta^{\mu\mu}/T^4$ is shallower for the asqtad action and shows a
%smaller cutoff dependence than results obtained with the p4 action.
In the transition region the cutoff effects can be well accounted for by a
shift of the $N_\tau=6$ data towards lower temperatures. This reflects
the cutoff dependence of the transition temperature and may also subsume
residual cutoff dependencies of the zero temperature observables used
to determine the temperature scale in the transition region.

\begin{figure}
\begin{center}
\resizebox{0.475\textwidth}{!}{%
  \includegraphics{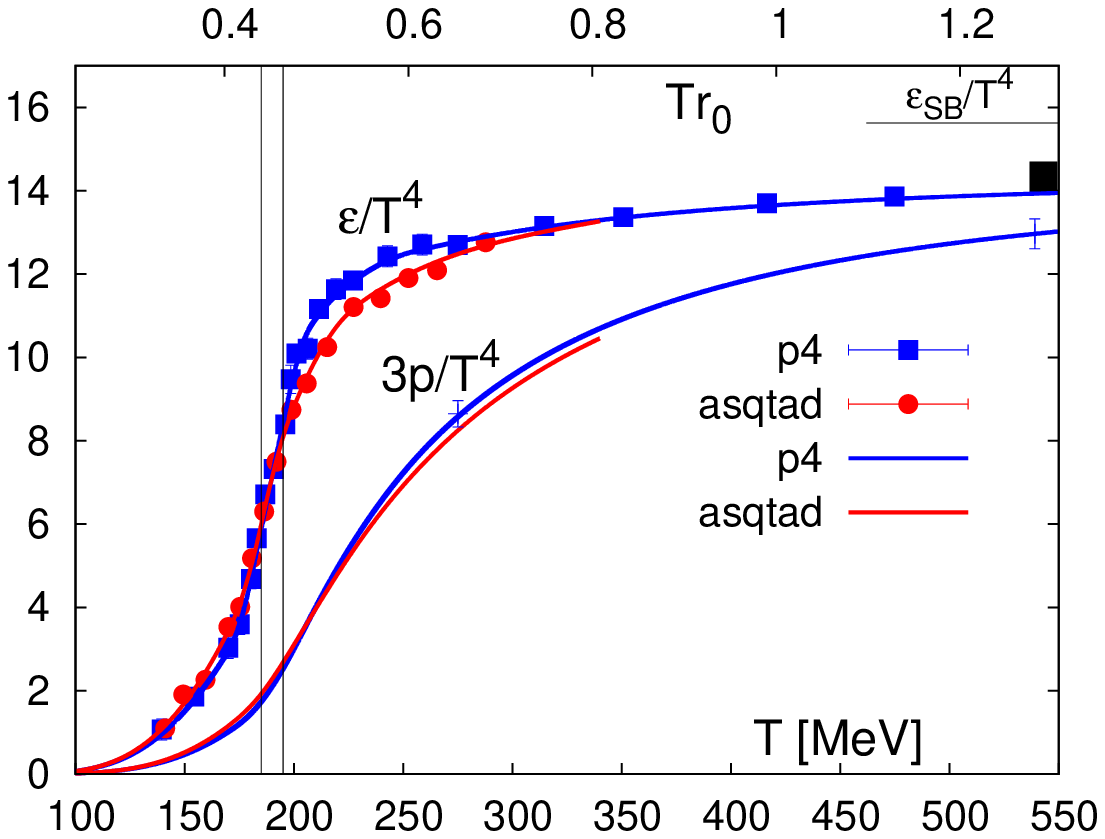}
}
\resizebox{0.475\textwidth}{!}{%
  \includegraphics{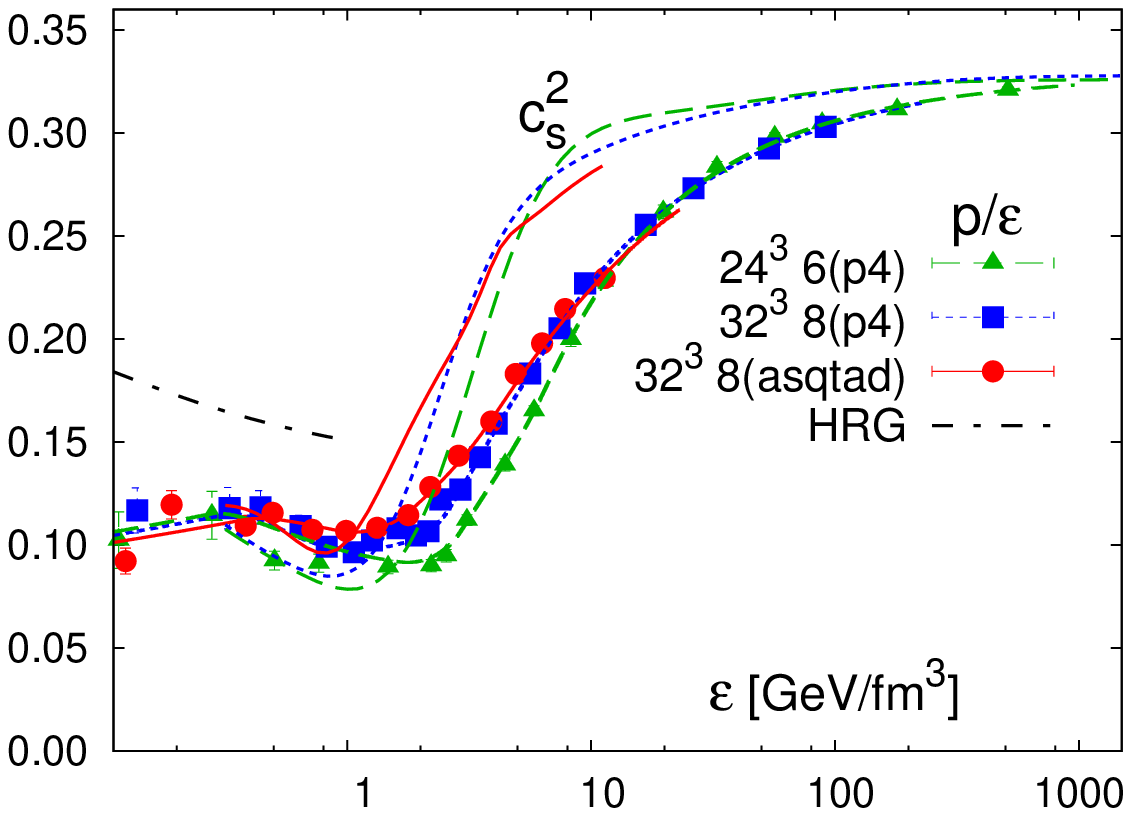}
}
\end{center}
\caption{
Energy density and three times the pressure as obtained by
p4 and asqtad actions on $N_\tau=8$ lattices (left) and pressure 
divided by energy density ($p/\epsilon$) and the square 
of the velocity of sound ($c_s^2$) (right), calculated on lattices with temporal extent 
$N_\tau =6,~8$. The $N_\tau =6$ results for the p4 action are taken
from Ref.~3.
\label{fig:e+3p}}
\end{figure}

In Fig.~\ref{fig:e+3p} we show results for the energy density and
three times the pressure.  Using Eq.~(\ref{delta}), the pressure is
obtained by integrating $\Theta^{\mu\mu}/T^5$ over the
temperature. Crosses with error bars indicate the systematic error on
the pressure that arises from different integration schemes.  The
energy density ($\epsilon$) is then obtained by combining results for
$p/T^4$ and $(\epsilon -3p)/T^4$. Finally, we show in in
Fig.~\ref{fig:e+3p}(right) $p/\epsilon$ as function of $\epsilon$. 
From a spline fit to that data
we also calculate the square of the velocity of sound, $c_s^2$.

\section{The QCD transition}
The bulk thermodynamic observables $p/T^4$, $\epsilon/T^4$ discussed
in the previous sections are sensitive to the change from hadronic to
quark-gluon degrees of freedom that occur during the QCD transition;
they thus reflect the deconfining features of this transition.  In a
similar vein, the temperature dependence of quark number
susceptibilities gives information on thermal fluctuations of the
degrees of freedom that carry a net number of light or strange quarks,
{\it i.e.}, $\chi_q \sim \langle N_q^2\rangle$, with $N_q$ denoting
the net number of quarks carrying the charge $q$. Quark number
susceptibilities change rapidly in the transition region as the
carriers of charge, strangeness or baryon number are heavy hadrons at
low temperatures but much lighter quarks at high temperatures.

\begin{figure}
\begin{center}
\resizebox{0.475\textwidth}{!}{%
  \includegraphics{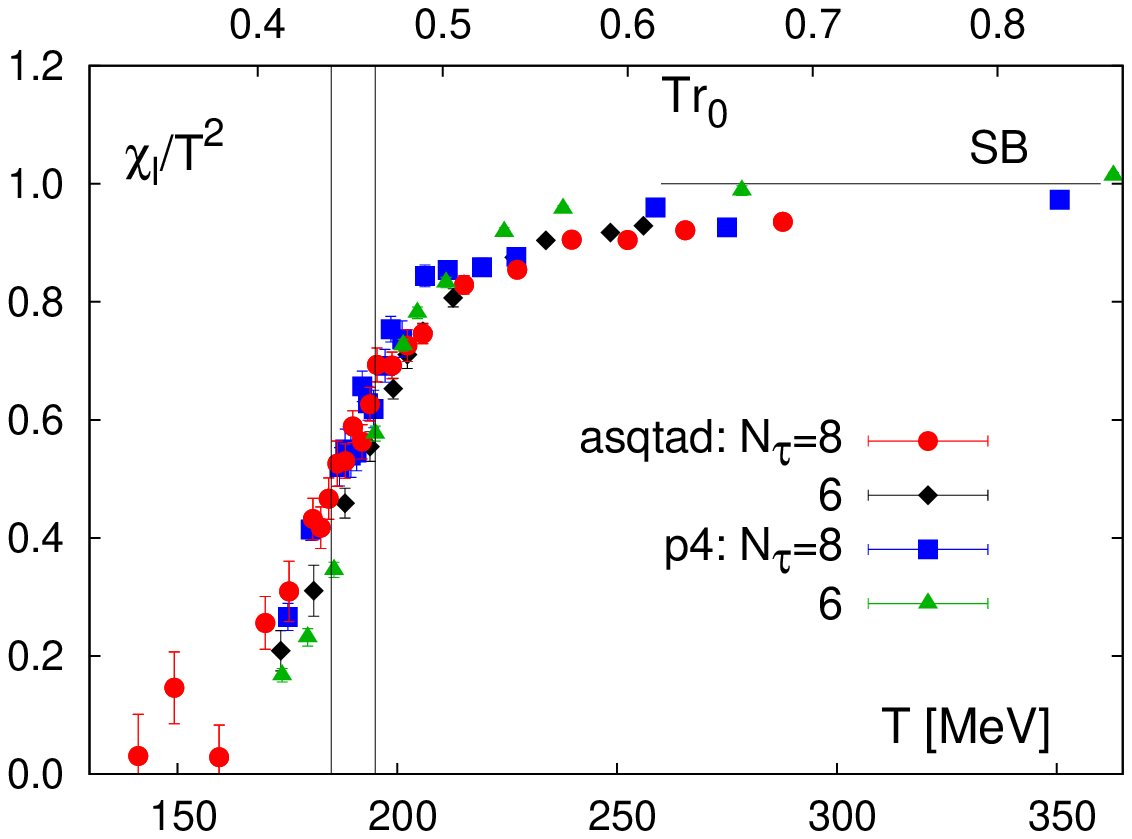}
}
\resizebox{0.475\textwidth}{!}{%
  \includegraphics{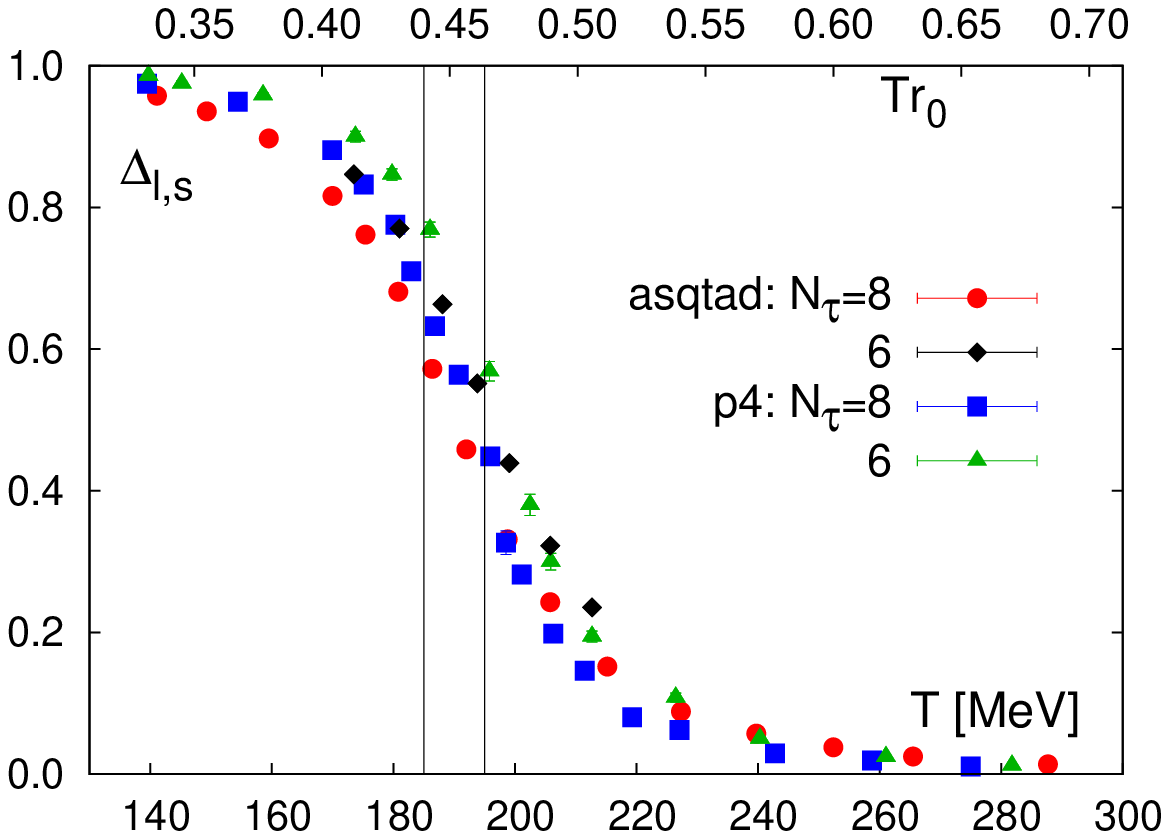}
}
\end{center}
\caption{
The light quark number susceptibility, calculated on lattices with temporal extent $N_\tau =6$ 
and $8$ (left) and the subtracted chiral condensate normalized to the corresponding 
zero temperature value (right). The band corresponds to a temperature 
interval $185\;{\rm MeV} \le T \le 195\;{\rm MeV}$. \label{fig:deco_chi}}
\end{figure}

Another important aspect of the QCD transition is the spontaneous breaking
(restauration) of the chiral symmetry. In Fig.~\ref{fig:deco_chi}(right)
we show the quantity $\Delta_{l,s}$, 
\begin{equation}
\Delta_{l,s}(T) = \frac{\langle \bar{\psi}\psi \rangle_{l,T} -
\frac{m_l}{m_s}
\langle \bar{\psi}\psi \rangle_{s,T}}{\langle \bar{\psi}\psi \rangle_{l,0} -
\frac{m_l}{m_s} \langle \bar{\psi}\psi \rangle_{s,0}} \; ,
\label{delta_ls}
\end{equation}
which is defined in terms of light and strange quark chiral condensates evaluated
at zero and non-zero temperature, respectively.
This is an appropriate observable that takes care of the additive
renormalizations in the chiral condensate by subtracting a fraction,
proportional to $m_l/m_s$, of the strange quark condensate from the
light quark condensate.  To remove the multiplicative renormalization
factor we divide this difference at finite temperature by the
corresponding zero temperature difference, calculated at the same
value of the lattice cutoff.

It is evident from Fig.~\ref{fig:deco_chi}, that $\Delta_{l,s}(T)$
varies rapidly in the same narrow temperature range as the bulk
thermodynamic observables and in particular the light quark number
fluctuations as shown in Fig.~\ref{fig:deco_chi}(left).  Based on this
agreement we conclude that the onset of liberation of light quark and
gluon degrees of freedom (deconfinement) and chiral symmetry
restoration occur in the same temperature range in QCD with almost
physical values of the quark masses, {\it i.e.}, in a region of the
QCD phase diagram where the transition is not a true phase transition
but rather a rapid crossover.

Furthermore, we note that the observed cutoff effects in the chiral
condensate can to a large extent be absorbed in a common shift of the
temperature scale.  A global shift of the temperature scale used for
the $N_\tau=6$ data sets by $5$~MeV for the p4 and by $7$~MeV for the
asqtad action makes the $N_\tau=6$ and $8$ data sets coincide almost
perfectly.  This is similar in magnitude to the cutoff dependence
observed in $(\epsilon-3p)/T^4$ and again seems to reflect the cutoff
dependence of the transition temperature as well as residual cutoff
dependencies of the zero temperature observables used to determine the
temperature scale.

\section*{Acknowledgment}
The work of CS has been supported by contract DE-AC02-98CH10886,
with the U.S. Department of Energy, and the Gesellschaft f\"ur
Schwerionenforschung under grant BILAER. Numerical simulations have been
performed on BlueGene/L computers at Lawrence Livermore
National Laboratory (LLNL) and the New York Center for Computational
Sciences (NYCCS).

\end{document}